\pdfoutput=1
\documentclass[aps, prl, reprint, tightenlines, superscriptaddress]{revtex4-1}
\usepackage{graphicx}
\usepackage[thickqspace]{SIunits}

\usepackage{color}
\usepackage{colortbl}
\newcommand{\mum}{\micro\meter}

\begin{document}
\title{Microwave guiding of electrons on a chip}

\author{J. Hoffrogge}
\affiliation{Max-Planck-Institut f\"ur Quantenoptik \\\small Hans-Kopfermann-Str.~1, D-85748
  Garching bei M\"unchen, Germany}
\author{R. Fr\"ohlich}
\affiliation{Max-Planck-Institut f\"ur Quantenoptik \\\small Hans-Kopfermann-Str.~1, D-85748
  Garching bei M\"unchen, Germany}
\author{M. A. Kasevich}
\affiliation{Department of Physics, Stanford University, Stanford, CA 94305, USA}
\author{P. Hommelhoff}
\altaffiliation{Corresponding author}
\affiliation{Max-Planck-Institut f\"ur Quantenoptik \\\small Hans-Kopfermann-Str.~1, D-85748
  Garching bei M\"unchen, Germany}

\date{\today}

\maketitle

{\bf  Electrons travelling in free space have allowed to explore fundamental physics like the wave nature of matter~\cite{Davisson1927, Hasselbach2010}, the Aharonov-Bohm~\cite{Chambers1960:AB, Tonomura1986:AB} and the Hanbury Brown-Twiss effect~\cite{Kiesel2002}. Complementarily, the precise control over the external degrees of freedom of electrons has proven pivotal for wholly new types of experiments such as high precision measurements of the electron's mass~\cite{Farnham1995} and magnetic moment~\cite{VanDyck1987, Hanneke2008} in Penning traps. Interestingly, the confinement of electrons in the purely electric field of an alternating quadrupole~\cite{Paul1990} has rarely been considered. Recent advances in the development of planar chip-based ion traps~\cite{Chiaverini2005, Seidelin2006, Leibrandt2009} suggest that this technology can be applied to enable entirely new experiments with electron beams guided in versatile potentials. Here we demonstrate the transverse confinement of a low energy electron beam in a linear quadrupole guide based on microstructured planar electrodes and driven at microwave frequencies. A new guided matter-wave system will result, with applications ranging from electron interferometry to novel non-invasive electron microscopy.}

Furthermore, together with advanced electron sources it appears feasible to prepare and guide electrons in the transverse motional ground state
in close analogy to light guided in single-mode optical fibres, as we discuss at the end of this letter. Appropriately structuring the guide will allow the (coherent) splitting and recombination of an electron beam as needed in matter-wave interferometry experiments.

\begin{figure}
\includegraphics[width=\columnwidth]{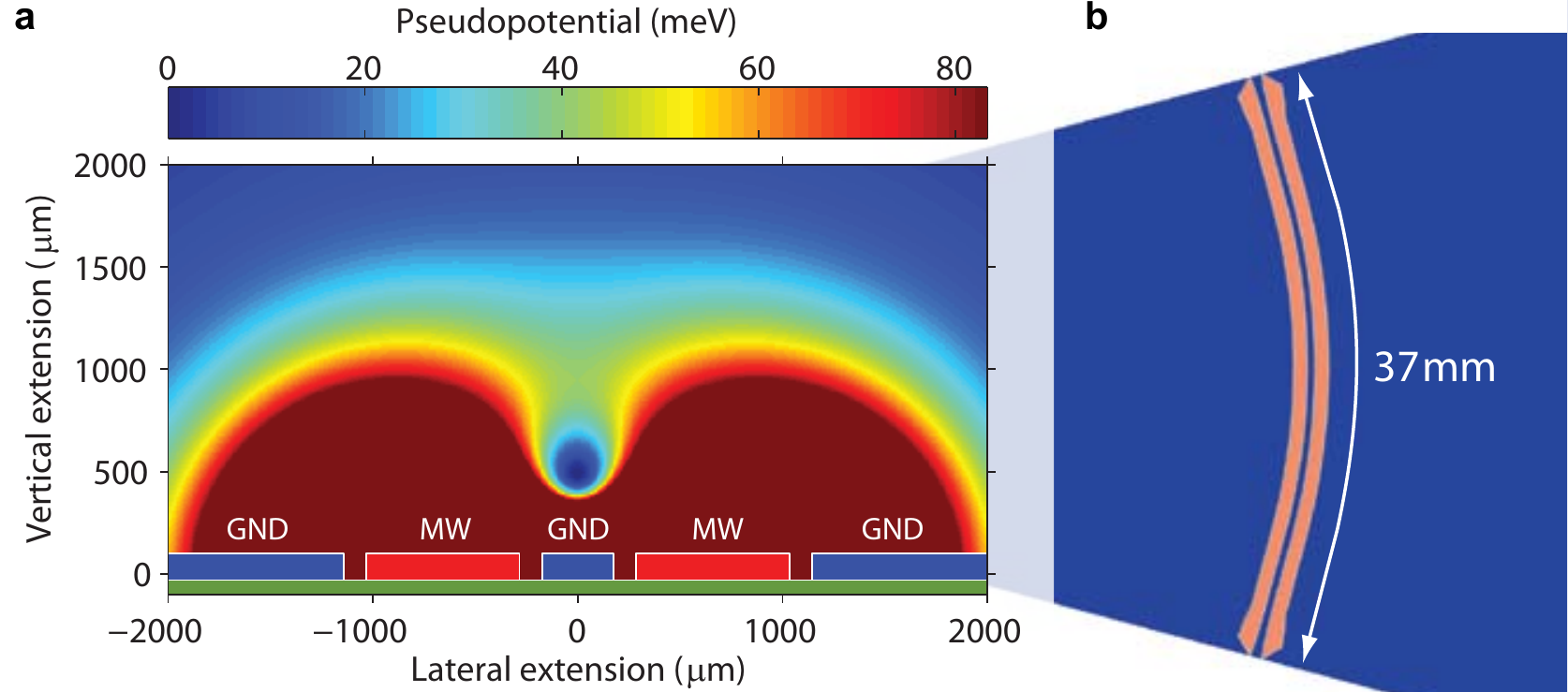}
	\caption{{\bf Pseudopotential and guide layout.} {\bf a}  Cut through the electrode plane with the pseudopotential experienced by an electron. The width of the electrodes is $350\,\mum$ for the centre one and $750\,\mum$ for the two adjacent. The outermost electrodes extend to the substrate edges. The gaps between the electrodes measure $110\,\mum$ while the electrode thickness is $40\,\mum$. The plotted height of the electrodes is exaggerated for illustration purposes. Guiding is achieved in the potential
minimum (blue) at a distance of $R = 500\,\mum$ above the central electrode. The potential is calculated for a driving frequency of $\Omega = 2\pi \cdot 970\,$MHz and the maximum peak voltage of $V = 33\,$V available in the experiment. Near the guide centre, the potential is approximately harmonic with a transverse trapping frequency of $\omega = 2\pi \cdot 133\,$MHz. Its depth is limited to $U = 41\,$meV by a saddle point forming above the centre. {\bf b} Electrode layout of the guiding structure. Electrons are guided along a bent five-wire-structure and deflected by $30\degree$.
The microwave signal is applied to the red electrodes, whereas the blue regions are grounded.}
	\label{fig:potential}
\end{figure}

To these ends it is highly desirable to shape the confining electromagnetic potential on small length scales. This can typically be done on the order of the distance between the trap centre and the field-generating electrodes. Hence, miniaturized traps with micro-structured electrodes allow for small and complex geometries. These have enabled quantum manipulation experiments both with neutral atoms in magnetic chip-traps~\cite{Riedel2010} and with ions in Paul traps~\cite{Brown2010, Harlander2010_2_Antenna, Blatt2008}. In analogy, microstructured Penning traps, combining a static magnetic field with the electric field generated by a planar electrode geometry, have been demonstrated for the three-dimensional confinement of electrons~\cite{Bushev2008}. To avoid the rather complicated dynamics in a magnetic field, we guide a propagating electron beam by means of a purely electric alternating quadrupole field.  This has so far only been realized with macroscopic structures~\cite{Weibel1961}, which impedes shaping the potential on a microscopic scale. We show in this letter that a planar electrode configuration is, besides its potential to generate complex waveguiding elements, an ideal choice to realize an electron guide as it is compatible with planar microwave transmission line technology to feed the structure.


The confinement of charged particles in a linear radiofrequency guide relies on the time-averaged action of an alternating electric field $E(\mathbf{r},t) = E(\mathbf{r}) \cos(\Omega t)$~\cite{Paul1990, Major_Werth2005}. In the ideal case, $E(\mathbf{r})$ is a pure quadrupole field generated by applying an alternating voltage with amplitude $V$ to electrodes at a distance $R$ from the guide centre. Particles can be confined if their transverse motion is slow compared to the drive frequency $\Omega$, which is quantified by a dimensionless stability parameter $q = 2 Q /m \cdot V / (\Omega^2 R^2)$ with $Q/m$ the charge-to-mass ratio of the particles. Stable transverse confinement is provided for $0 < q < 0.91$. For small $q$, the particle's transverse motion can be approximated by that in a harmonic pseudopotential with frequency $\omega = q/\sqrt{8}\cdot \Omega$ and depth $U = q/8\cdot V$.

Compared to the confinement of ions, electrons with their $\sim 10^{4}$ times higher $Q/m$ require notedly different driving parameters to keep $q$ small and thus the guide stable. Here, we employ a driving frequency of $\Omega \approx 2 \pi \cdot 1\,$GHz with a structure size of $R = 500\,\mum$. This keeps $V$ in manageable limits, $V \approx 30\,$V. The driving wavelength $\lambda$ is still much larger than the longitudinal structure size $L$ ($\lambda = 21\,$cm vs. $L = 37\,$mm), which allowed us to work in a standing-wave configuration.


The guiding field $E(\mathbf{r},t)$ is generated by applying the drive voltage to a set of five electrodes on a planar substrate, in close analogy to planar ion traps~\cite{Chiaverini2005, Pearson2006, Seidelin2006}. In Figure~\ref{fig:potential}a a cut through the electrode layout is shown together with the microwave pseudopotential experienced by an electron above the substrate. For the microwave power available in this experiment (10\,W) we are able to realize transverse trapping frequencies up to $\omega = 2\pi\cdot 133\,$MHz and potential depths up to $U = 41 \,$meV.

The complete electrode layout of the guide is shown in Figure~\ref{fig:potential}b. It consists of a 37\,mm long curve with a bending radius of 40\,mm spanning an angle of $30\degree$. Electrons are injected at one end of the structure, travel along the curve and are ejected at the other end, where they are detected by an imaging micro-channel plate. In such a guide an electron with a kinetic energy of $E_{\mathrm{kin}} = 2\,$eV, as typically used here, experiences around four oscillations in the pseudopotential, corresponding to 44 oscillations of the driving field, while travelling along the guide. Near the edges of the substrate we have optimized the wire shape numerically in order to achieve smooth coupling into the guide (see Supplementary Information). The microwave signal is fed to the electrodes by a coplanar transmission line on the bottom side, see methods.

\begin{figure}
\includegraphics[width=\columnwidth]{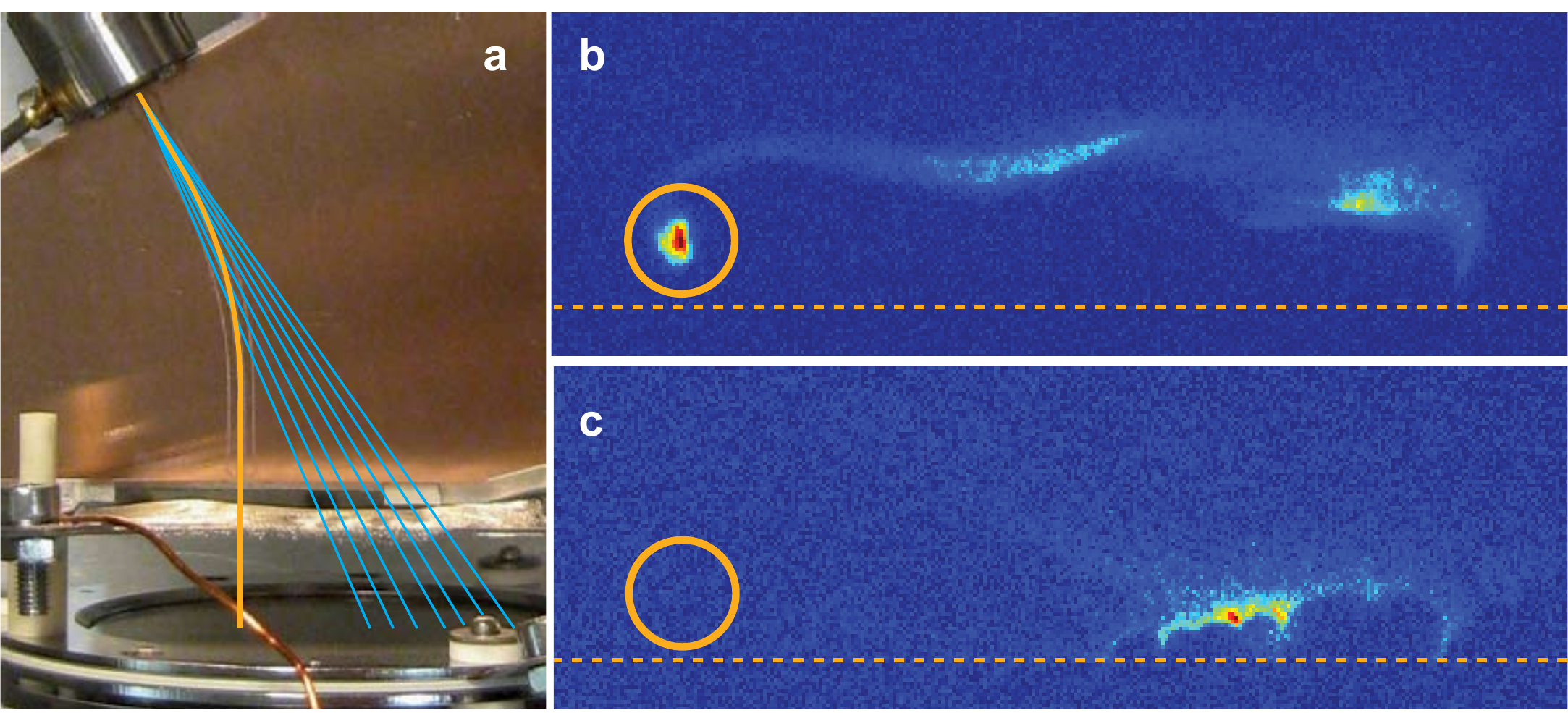}
\caption{{\bf Images of setup and guided electrons.} {\bf a}~ Experimental setup with the substrate in the centre. The electron gun is visible at the top left, the imaging microchannel plate electron detector (MCP) at the bottom. Guided electrons follow the orange curve from source to MCP, whereas trajectories of unguided electrons are indicated in blue.  {\bf b}~MCP image. The orange circle indicates the position of the guide's exit port, the horizontal dashed line the electrode plane. A bright spot of guided electrons is clearly visible at the exit. The smaller signal in the centre and to the right is caused by electrons lost during guiding, see text. The guide is operated at $\omega = 2\pi\cdot 100\,$MHz, $U = 27\,$meV and $q = 0.3$. Recorded data look similar over the whole parameter range (up to $\omega = 2\pi\cdot 133\,$MHz and $U = 41\,$meV) accessbile with the microwave power available in the experiment. {\bf c}~For comparison: image with no microwave power applied to the guide. Electrons fanning out from the source are visible as broadly distributed signal to the right. Its structure can be attributed to charging of the substrate exposed between the electrodes, see text for details. Note that the image intensity has been increased by a factor of two as compared to {\bf b}.
The kinetic energy of the elctrons in both {\bf b} and {\bf c} is $4\,$eV.}
\label{fig:guiding}
\end{figure}

Figure~\ref{fig:guiding}a shows a photograph of the experimental setup with trajectories of guided and unguided electrons indicated by orange and blue lines, respectively. Guiding is demonstrated by forcing the confined electrons on a curved path that ends on the left side of the detector when watched from behind.
For appropriate settings of $\Omega$ and $V$ we obtain a bright spot of electrons visible exactly at the position of the guide exit (Figure~\ref{fig:guiding}b).

Due to imperfect coupling and charging of the substrate between the electrodes a part of the electrons is lost while passing over the substrate. Electrons lost in the bent section are visible as faint curved horizontal line in the image centre, whereas those lost at the beginning of the guide form a brighter spot on the right-hand side. The fraction of guided electrons is expected to significantly improve with a better collimated source and future electroplated substrates with a high aspect ratio of electrode thickness over gap width that will shield the guided electrons from exposed substrate areas~\cite{Chiaverini2005, Amini2008}.

For comparison Figure~\ref{fig:guiding}c shows a detector image without microwave power applied. In that case only a diffuse spot of electrons fanning out from the gun is visible to the right of the guide's exit port.
The dark regions between electrons and substrate indicate that electrons are deflected away from the gaps between the electrodes due to substrate charging. Its effect can partially be compensated by applying a voltage of up to several volts to a plate $10\,$mm above the substrate, leading to robust electron guiding for kinetic energies from approximately $1\,$eV to $5\,$eV.


\begin{figure}%
\includegraphics[width=\columnwidth]{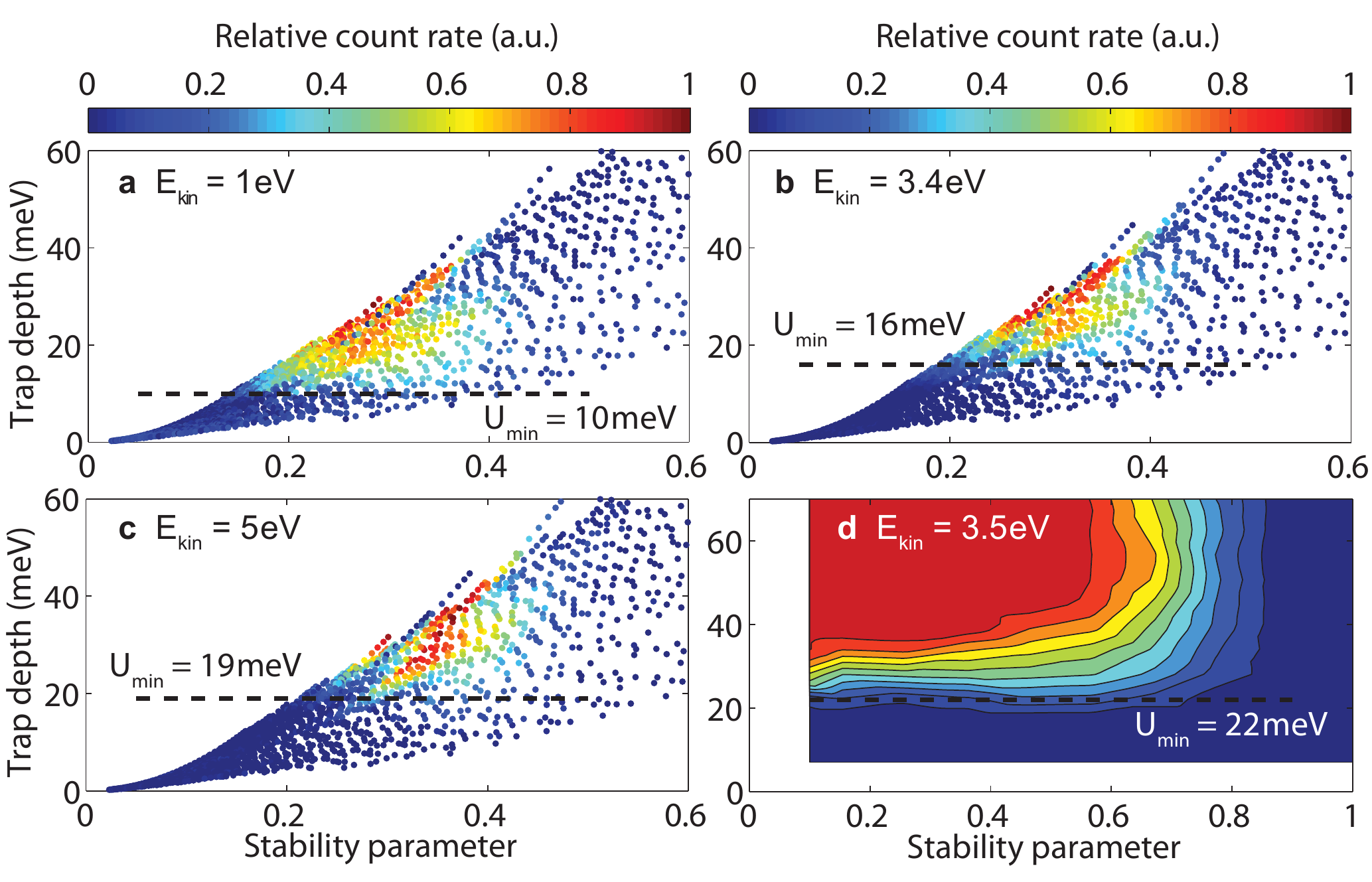}%
\caption{{\bf Fraction of guided electrons as a function of stability and potential depth.} {\bf a~-~c}~Experimental data for different electron energies as indicated. With increasing kinetic energy the minimum potential depth $U_{\mathrm{min}}$ needed for guiding increases. Also, the guide becomes unstable for stability parameters higher than approximately $0.42$. As the gain of the MCP and the beam current emanating from the electron gun vary from {\bf a} to {\bf c}, each plot has been normalized separately. The white areas in the upper left half of the plots were not accessible for technical reasons (limited microwave power). {\bf d}~Results of a particle tracking simulation at $3.5\,$eV kinetic energy. Note the different plot range of the horizontal axis. See text and Supplementary Information for details.}%
\label{fig:stabis}%
\end{figure}

For a thorough guide characterization we have recorded the number of guided electrons as a function of potential depth $U$ and stability parameter $q$ (Figure~\ref{fig:stabis}, raw data in terms of $V$ and $\Omega$ are presented in the Supplementary Information). It is apparent that a minimum potential depth $U_{\mathrm{min}}$ is necessary to counter the centrifugal force on the particles in the curved guide.
As expected, when $E_{\mathrm{kin}}$ is increased $U_{\mathrm{min}}$ also increases due to the larger centrifugal force (from $U_{\mathrm{min}} \approx 10\,$meV at 1\,eV to $U_{\mathrm{min}} \approx 19\,$meV at 5\,eV).
Furthermore, for $U > U_{\mathrm{min}}$ we observe a constant signal of guided electrons up to $q \approx 0.42$. The loss of guiding for higher $q$ can be attributed to radiofrequency heating as the micromotion of the electrons in the guide is increased.


Figure~\ref{fig:stabis}d shows the result of a numerical particle-tracking simulation of electron trajectories in the alternating field, see Supplementary Information. In qualitative agreement with the experimental data, we observe a loss of guiding below $U_{\mathrm{min}} \approx 22\,$meV and above $q \approx 0.8$ for $E_{\mathrm{kin}} = 3.5\,$eV. We attribute the differences to the experimentally recorded values to calibration errors of the microwave power fed to the electrodes, a larger transverse electron momentum in the experiment and numerical uncertainties in the simulation. The particle-tracking results also confirm that guiding should work down to $q$ approaching 0, where we have not been able to record data due to limited microwave power available.


The next important experimental step lies in the realization of a beam splitter similar to that developed for ions~\cite{Amini2010}. Furthermore, scaling to a guide-to-surface distance of $R = 50\,\mum$ and a driving frequency of $\Omega = 2\pi\cdot 10\,$GHz would result in a system providing even faster dynamics at a transverse trapping frequency of $\sim$$1.2\,$GHz.
The possibility of lithographic substrate patterning will allow to extend the guiding structures to more complex geometries and to many electrons guided in separate potentials.

A fascinating possibility for a wholly new quantum device arises from the combination of an electron guide as demonstrated here with a single-atom tip electron emitter~\cite{Fink88, Hasselbach2010}. These sources have been shown to emit electrons fully coherently~\cite{Chang2009} and are Heisenberg-uncertainty limited in terms of emitter source size and transverse electron momentum. The same minimum uncertainty criterion also applies to position and momentum of the ground state wavefunction in a static harmonic potential, which describes the time-averaged transverse motion of electrons near the centre of the guide to very good approximation. Thus, with electron optics preserving phase space density it should be possible to directly prepare electrons originating from a single-atom tip in the transverse ground state of the guiding potential {\em without} the need for cooling (note that the ground state wavefunction of the time-averaged pseudopotential is still physically meaningful  in the real, time-varying potential~\cite{Leibfried2003_RMP}). This represents an ideal starting point for guided matter-wave interferometry~\cite{Hasselbach2010, Andersson02} and quantum manipulation experiments.

An electron guide together with femtosecond laser-triggered sources~\cite{Hommelhoff_PRL2006} will enable full 4-d control of electrons. Hence one can envision controlled-interaction experiments between, for example, two  electrons propagating in neighbouring guides and interacting via the Coulomb force, closely related to what has very recently been demonstrated with ions~\cite{Brown2010, Harlander2010_2_Antenna}. Because of the smaller mass of electrons, comparable coupling strengths of $\Omega_c \approx 2\pi\cdot 1\,$kHz can be achieved over a ten times larger distance of $500\,\mum$ (assuming transverse coupling in adjacent guides with $\omega = 2\pi\cdot 100\,$MHz). With $\omega = 1\,$GHz and $R = 50\,\mum$ the coupling strength increases to $\Omega_c \approx 100\,$kHz. One can also envision interfacing guided electrons with other quantum systems like trapped ions, atoms, or electrons in solids. With an additional potential the guide can be longitudinally closed and converted to a 3-d trap. Furthermore, it has recently been proposed to use laterally confined electrons in a microstructured potential for novel non-invasive electron microscopy~\cite{Putnam2009}.

Technically, the high transverse frequencies will well isolate the electrons from electric field noise. The heating rate $\dot{n}$ in quanta per second is given by $\dot{n} = e^2 / \left(4m\hbar\omega \right) \cdot S_E(\omega)$ with the electric noise density $S_E(\omega) \propto 1/\omega \cdot 1/R^{-4}$~\cite{Labaziewicz2008}. Scaling the results measured for ions in microscopic traps at room temperature to electrons in the guide demonstrated here yields a heating rate of $\dot{n} \approx 30 /$s. Thus a ground state electron can perform around 60 million oscillations in the pseudopotential before being heated to the first excited state.

\section{Methods}

{\small
The guide is fabricated on a Rogers RO4350B microwave substrate with electrodes made from gold plated copper with $40\,\mum$ thickness. The microwave signal is fed to the substrate via an edge-mount SMA connector. On the substrate, the signal is conducted by a coplanar transmission line on the bottom side of the substrate, which runs perpendicular to the guide on the top side and is connected by vias of $150\,\mum$ diameter to the centre of the guide. Because the electrodes' ends are open, a standing wave forms with antinodes at the beginning and the end of the guide. Due to the small length of the whole structure the voltage difference along the guide is measured to be less than 10\,\%.

The microwave signal is directly fed from an amplifier to the substrate without using any resonating structures. It is generated by an Agilent E8257C generator, boosted in a Minicircuits ZHL-30W-252-S amplifier up to 30\,W, and sent via an SMA-feedthrough into the vacuum chamber. Inside the chamber, a 40\,cm long vacuum compatible co-axial cable connects the feedthrough to the substrate. With a bias-tee (MECA 200S-FF-2) between amplifier and feedthrough static charging of the signal conductors is prevented. An additional directional coupler (MECA 780-20-9.700) allows to monitor the microwave power fed to the substrate. The power applied to the electrodes is inferred from this signal by correcting for the independently measured frequency-dependent loss of the microwave cables and the transmission line structure on the substrate.

The electron gun consists of a thermal source and beam forming elements~\cite{Erdman1982} with an exit aperture of $20\,\mum$ diameter. Typical electron energies used here range from 1\,eV to 10\,eV with typical beam currents on the order of several ten nanoamperes. The gun is mounted on a three-axis translation stage in order to position the aperture precisely in front of the guide. The guide is shielded from electric fields by a metallic cover with its top plate 10\,mm above the substrate (which has been removed to take the picture shown in Figure~\ref{fig:guiding}a).
The guiding region is additionally shielded against electric fields emanating from the MCP by a grounded mesh between substrate and MCP. The whole setup is surrounded by a mu-metal box to reduce magnetic stray fields to a level of $B = 6\,$mG and installed in a vacuum chamber held at a pressure of $1\cdot 10^{-6}\,$mbar.}

Correspondence and requests for materials should be addressed to P.H.
(email: peter.hommelhoff@mpq.mpg.de).

\section{Acknowledgement} We thank Ph. Treutlein for discussions and making available to us his code for simulating transmission lines and M. Herrmann for discussions and help simulating electron trajectories in the guide. We thank W. H{\"a}nsel for discussions. We also thank the H\"ansch group for lending us equipment.

\section{Author Contributions}

Experimental work by J.H. and R.F.; theoretical work by P.H., J.H. and M.A.K.

\section{Competing Financial Interest}
The authors declare that they have no competing financial interest.

\section{Supplementary Information}





\subsection{Electron Coupling Structure}

\begin{figure}
\includegraphics[width=\columnwidth]{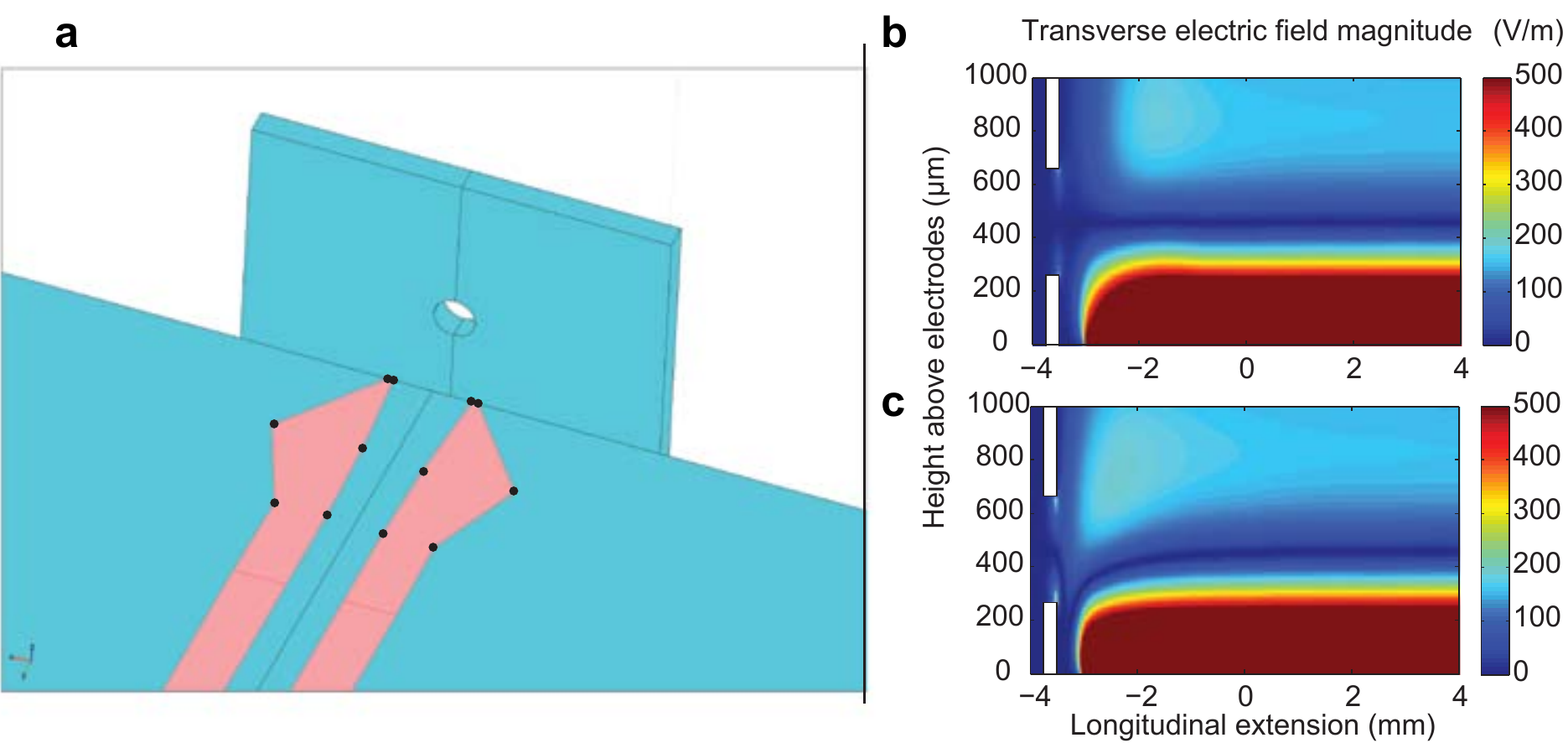}%
	\caption{{\bf Coupling structure.} {\bf a} Electrode pattern at the edges of the substrate with aperture in the back. Signal conductors are shown in pink, grounded structures in turquoise. In order to enable smooth coupling to the guide the shape of the electrodes has been numerically optimized by lateral variation of twelve points at the edges of the signal electrodes, as indicated by black dots. {\bf b} Optimization result. Finite element simulation of the magnitude of the radial electric field in a vertical cut through the guide centre at the beginning of the structure. After optimization, the guiding minimum extends horizontally through the aperture. {\bf c} For comparison, with electrodes ending in straight lines, the line of minimum radial field is bent down between the end of the guide and the coupling aperture in front.} 
	\label{fig:design}%
	\end{figure}
	
Electrons are injected into the guide at the edges of the substrate. Here, the electrode pattern has been optimized numerically~\cite{Amini2010} to achieve a smooth extension of the guide's potential minimum into the field free region in front of the substrate. We calculated the electric field above the electrodes using the elctrostatic field solver of the Comsol package. This static approximation is justified for structures much smaller than the wavelength of the driving field, as it is here the case.
We used Matlab's built-in Nelder-Mead Simplex algorithm for optimization. In the calculation we varied the position of twelve points, shown in Figure~\ref{fig:design}a, at the edges of the signal conductors laterally while preserving symmetry with respect to the vertical plane through the guide centre. The optimization goal was to minimize the magnitude of the transverse electric field along the guide axis. We have also included an aperture plate placed at a distance of $500\,\mum$ in front of the guide to account for the last element of the electron gun.
The result of this optimization is presented in Figure~\ref{fig:design}b where the magnitude of the transverse electric field is shown in a vertical plane orientated along the guide and cut through the potential minimum. Here, the potential minimum extends in a horizontal line from the guide through the aperture (white rectangles). The maximum field on the guide axis has been reduced by the optimization to $E_{\mathrm{max}} = 6\,$V/m, a factor 20 lower than for a geometry ending in straight wires. For comparison, the field of this straight configuration is plotted in Figure~\ref{fig:design}c. The potential minimum bends down between the substrate and the aperture plate leading to a maximum transverse field of $E_{\mathrm{max}} = 120\,$\,V/m on the trap axis.

\subsection{Stability Measurements}

\begin{figure}
\includegraphics[width=\columnwidth]{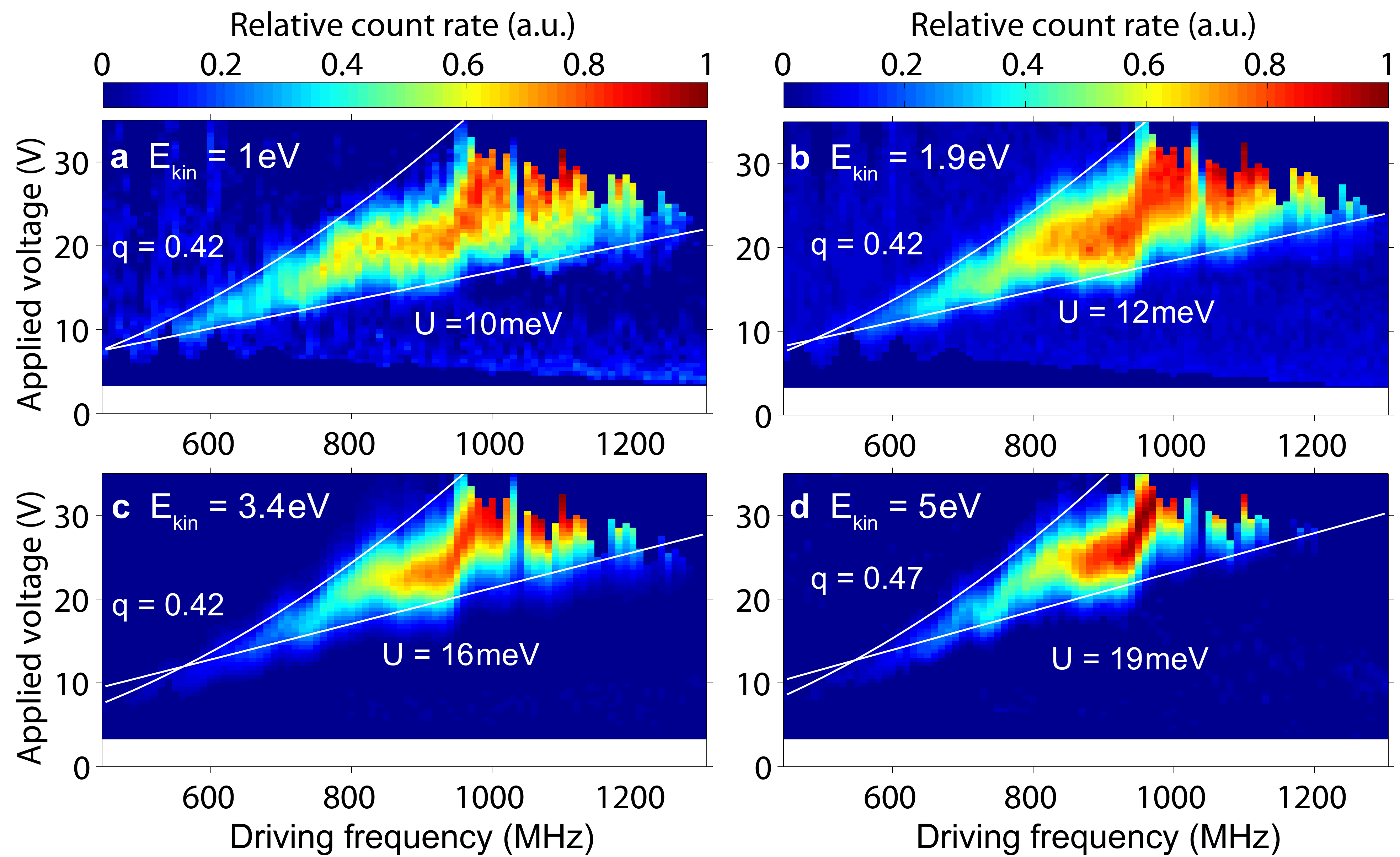}
	\caption{{\bf Electron count rate as function of the driving parameters for different electron energies.} {\bf a-d}  Relative count rate of guided electrons for varying driving frequency $\Omega$ and voltage amplitude $V$. The white lines indicate parameter settings whith constant stability parameter $q$ and potential depth $U$ respectively. For each electron energy displayed, we obtain guiding below a maximum $q$ and above a minimum $U$. We attribute the drop in count rate for all energies at 1040\,MHz driving frequency to technical reasons.}%
	\label{fig:stabis_fV}%
\end{figure}

In order to characterize the stability of the electron guide we recorded the number of guided electrons for a range of driving frequencies $\Omega$ and voltage amplitudes $V$. Figure~\ref{fig:stabis_fV} shows four such measurements for four different electron energies. Each plot has been taken with a different MCP gain setting to compensate for the decreasing emission current of the electron gun with decreasing electron energy. The data for each kinetic energy have been normalized to the peak count rate at that energy separately. The voltage axis is derived from measurements of the microwave amplifier output power, the loss in the circuitry connecting the substrate (approx. $1\,$dBm) and the loss on the substrate itself (approx. $2\,$dBm to $5\,$dBm, depending on frequency).

For illustration purposes we have added lines of constant stability parameter $q$ and potential depth $U$ that bound the electron signal from above and below. These have been derived from the driving parameters by
\[
q = \sqrt{8} \frac{\omega}{\Omega} = \eta\frac{2Q}{m} \frac{V}{\Omega^2 R^2}
\]
and 
\[
U = u\frac{Q^2}{\sqrt{4}m} \frac{V^2}{\Omega^2 R^2}.
 \]
The dimensionless parameter $\eta$ quantifies the reduction in transverse frequency compared to a perfect quadrupole field ($\eta = 1$) and has been determined from the numerically calculated pseudopotential (see Figure~1a of the main text) above the guide to $\eta = 0.31$.
Likewise, $u$ quantifies the reduction in potential depth with $u = 0.0079$. These equations have also been used to generate the plots displayed in Figure~3a-c of the main text. As the microwave power available was limited by the maximum output power of the amplifier, we were not able to record data for voltages above $V \approx 30\,$V. This explains the lack of data for small $q$ and larger $U$ in Figure~3a-c.

\begin{figure}%
\includegraphics[width=\columnwidth]{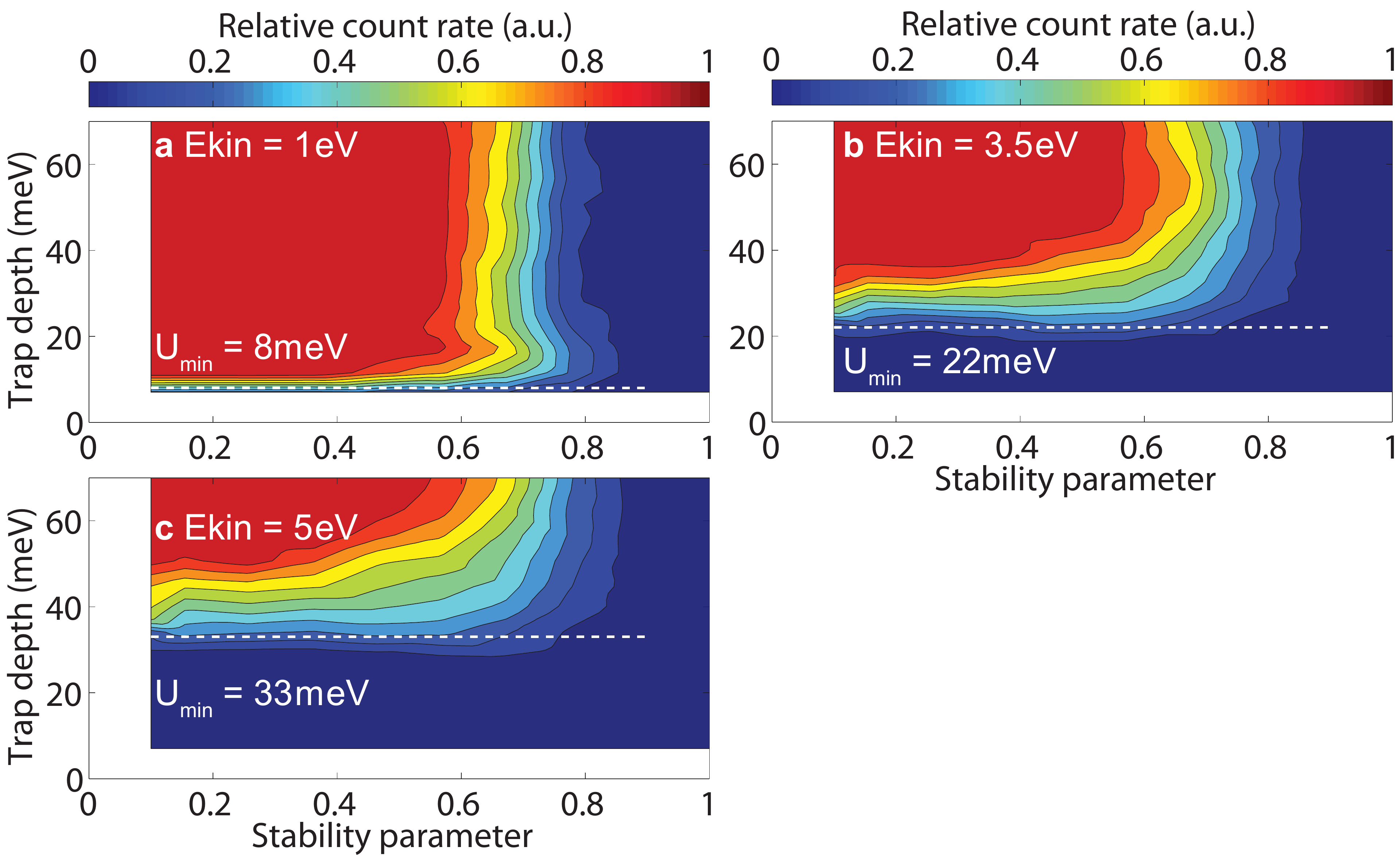}
\caption{{\bf Simulated electron count rate versus guide stability and depth.} The minimum potential depth needed for guiding increases with the electrons kinetic energy, as also experimentally observed. Guiding is lost for stability parameters larger than $q \approx 0.8$. The calculation also confirms that guiding is preserved for $q$ approaching zero.}%
\label{fig:stabisSim}%
\end{figure}

\subsection{Simulation of Electron Trajectories}

As a comparison to the measurements of the dynamics of the microwave guide presented in Figure 3a-c of the main text, we performed a numerical simulation of electron trajectories in the guiding potential using the commercial CPO-package. The program first solves for the electrostatic charge configuration on the guide electrodes and then adds a periodic time dependence before tracking the particles. This quasistatic approach is justified for electrically small structures with dimensions large compared to the driving wavelength as in this experiment. Besides the bent electrode structre we also included a grounded aperture in front of the guide, similar to that shown in Figure~\ref{fig:design}.

For each pair of electrode voltage $V$ and driving frequency $\Omega$ we simulated 25 trajectories forming the envelope of a beam starting 1.5\,mm behind the aperture from a circular disk with $20\,\mum$ diameter and approximately $1\degree$ full divergence angle. In addition, we averaged over 16 different initial phases of the guiding field. At the exit of the guide the number of electrons passing a circular disc of $100\,\mum$ radius around the guide axis has been recorded.
The results of the simulations are shown in Figure 3c in the main text and additionally in Figure~\ref{fig:stabisSim}. The simulation yields the expected dependence on potential depth $U$: A minimum $U$ is needed to compensate for the centrifugal force on the guided particles. This minimum depth increases from $U \approx 8\,$meV at 1\,eV kinetic energy to $U \approx 33\,$meV at 5\,eV. We attribute the differences between simulation and experiment to an imperfect calibration of losses in the microwave circuitry, a higher transverse electron momentum in the experiment and numerical uncertainties.
The simulation also confirms that guiding should work at constant potential depth down to $q = 0$, as expected from theory~\cite{Major_Werth2005}.

{~}

{~}





\end{document}